\documentclass[a4paper,fleqn,usenatbib]{mnras}

\usepackage{newtxtext,newtxmath}

\usepackage[T1]{fontenc}
\usepackage{ae,aecompl}
\usepackage{threeparttable}
\usepackage{hyperref}
\usepackage[dvipsnames]{xcolor}

\usepackage{graphicx}	
\usepackage{amsmath}	
\usepackage{amssymb}	

\newcommand{\zsol}{Z$_{\odot}$}

\newcommand{\lya}{Lyman-$\alpha$ }
\newcommand{\lyans}{Lyman-$\alpha$}
\newcommand{\oiii}{\hbox{[O\,{\sc iii}]}}
\newcommand{\hii}{\hbox{H\,{\sc ii}}}
\newcommand{\siii}{\hbox{[S\,{\sc iii}]}}
\newcommand{\sii}{\hbox{[S\,{\sc ii}]}}
\newcommand{\nii}{\hbox{[N\,{\sc ii}]}}
\newcommand{\xii}{$\xi_{\mathrm{ion}}$}
\newcommand{\fesc}{$f_{\mathrm{esc}}$}
\newcommand{\muv}{$M_{\mathrm{UV}}$}
\newcommand{\fion}{$f_{\mathrm{esc}}^{\mathrm{ion}}$}
\newcommand{\flya}{$f_{\mathrm{esc}}^{\mathrm{Ly\alpha}}$}
\newcommand{\ha}{\hbox{H $\alpha$}}
\newcommand{\hb}{\hbox{H $\beta$}}

\title[Elevated ionizing photon production efficiency in faint high-EW Lyman-$\alpha$ emitters]{Elevated ionizing photon production efficiency in faint high-equivalent-width Lyman-$\alpha$ emitters\thanks{Based on observations made with ESO telescopes at the La Silla Paranal Observatory under program IDs 094.A-2089(B), 095.A-0010(A), 096.A-0045(A), and 096.A-0045(B).}}

\author[Maseda et al.] {Michael V. Maseda$^{1}$\thanks{E-mail: maseda@strw.leidenuniv.nl}, 
Roland Bacon$^{2}$, 
Daniel Lam$^{1}$,
Jorryt Matthee$^{3}$, \newauthor
Jarle Brinchmann$^{4,1}$,
Joop Schaye$^{1}$, 
Ivo Labbe$^{5}$,
Kasper B. Schmidt$^{6}$,\newauthor
Leindert Boogaard$^{1}$, 
Rychard Bouwens$^{1}$,
Sebastiano Cantalupo$^{3}$,
Marijn Franx$^{1}$, \newauthor
Takuya Hashimoto$^{7}$,
Hanae Inami$^{8}$,
Haruka Kusakabe$^{9}$,
Guillaume Mahler$^{10}$,\newauthor
Themiya Nanayakkara$^{1}$,
Johan Richard$^{2}$,
and Lutz Wisotzki$^{6}$\\
  $^{1}$Leiden Observatory, Leiden University, P.O. Box 9513, 2300 RA, Leiden, The Netherlands \\
  $^{2}$Univ Lyon, Univ Lyon 1, CNRS, Centre de Recherche Astrophysique de Lyon UMR5574, F-69230, Saint-Genis-Laval, France\\
  $^{3}$ETH Z\"urich, Department of Physics, Wolfgang-Pauli-Str. 27, 8093 Z\"urich, Switzerland\\
  $^{4}$Instituto de Astrof{\'\i}sica e Ci{\^e}ncias do Espa\c{c}o, Universidade do Porto, CAUP, Rua das Estrelas, PT4150-762 Porto, Portugal\\
  $^{5}$Centre for Astrophysics and Supercomputing, Swinburne University of Technology, Hawthorn, Victoria 3122, Australia\\
    $^{6}$Leibniz-Institut f\"ur Astrophysik Potsdam (AIP), An der Sternwarte 16, 14482 Potsdam, Germany\\
  $^{7}$Faculty of Science and Engineering, Waseda University, 3-4-1 Okubo, Shinjuku, Tokyo 169-8555, Japan\\
  $^{8}$Hiroshima Astrophysical Science Center, Hiroshima University, 1-3-1 Kagamiyama, Higashi-Hiroshima, Hiroshima 739-8526,
Japan\\
  $^{9}$Observatoire de Gen\`eve, Universit\'e de Gen\`eve, 51 Ch. des Maillettes, 1290 Versoix, Switzerland\\
    $^{10}$Department of Astronomy, University of Michigan, 1085 South University Ave., Ann Arbor, Michigan 48109, USA}
\date{Accepted 2020 February 21. Received 2020 February 21; in original form 2019 October 25}

\pubyear{2020}

\begin{document}
\label{firstpage}
\pagerange{\pageref{firstpage}--\pageref{lastpage}}
\maketitle

\begin{abstract}
While low-luminosity galaxies dominate number counts at all redshifts, their contribution to cosmic Reionization is poorly understood due to a lack of knowledge of their physical properties.   We isolate a sample of 35 $z\approx4-5$ continuum-faint \lya emitters from deep VLT/MUSE spectroscopy and directly measure their \ha\ emission using stacked \textit{Spitzer}/IRAC Ch. 1 photometry.  Based on \textit{Hubble} Space Telescope imaging, we determine that the average UV continuum magnitude is fainter than $-$16 ($\approx$0.01 $L^{\star}$), implying a median \lya equivalent width of 249 \AA.  By combining the \ha\ measurement with the UV magnitude we determine the ionizing photon production efficiency, \xii, a first for such faint galaxies.  The measurement of log$_{10}$ (\xii\ [Hz erg$^{-1}$]) = 26.28 ($^{+0.28}_{-0.40}$) is in excess of literature measurements of both continuum- and emission line-selected samples, implying a more efficient production of ionizing photons in these lower-luminosity, \lyans-selected systems.  We conclude that this elevated efficiency can be explained by stellar populations with metallicities between 4$\times$10$^{-4}$ and 0.008, with light-weighted ages less than 3 Myr.
\end{abstract}


\section{INTRODUCTION}

Although recent observations have provided ever more certainty about the timing and duration of the last significant phase transition in the Universe, cosmic Reionization \cite[e.g][]{2018arXiv180706209P,2018Natur.553..473B}, much remains to be understood about the source(s) of the photons which caused the change.  Evidence is mounting that star-forming galaxies could have produced enough ionizing photons to drive Reionization at $z>6$, but typically only under the assumption that galaxies with UV magnitudes much fainter than the characteristic luminosity ($L^{\star}$) dominate the total number counts and that these galaxies have similar physical properties to brighter systems \cite[e.g.][]{2004ApJ...600L...1Y,2012ApJ...752L...5B,2012ApJ...758...93F,2013ApJ...768...71R}.  

While observed UV luminosity functions indicate that faint galaxies are numerous at all redshifts \cite[e.g.][]{2018MNRAS.479.5184A}, few direct observational (spectroscopic) constraints on the efficiency of their ionizing photon production exist.  This difficulty is primarily due to the observability of spectral features, particularly those in the rest-frame optical: ground-based spectroscopy can only cover features such as \ha\ in the near-IR until $z\approx$ 2.8.  \ha\ in particular is crucial to understand the contribution of galaxies to Reionization as it is directly related to the intrinsic production rate of ionizing photons (compared to the resonantly-scattered \lyans).  The ratio of the \ha\ flux to the flux of non-ionizing (UV) photons, when combined with the escape fraction of ionizing photons and the number density of galaxies, can be used to determine the total production rate of ionizing photons in the early Universe.

Although detecting \ha\ is currently not possible with traditional spectroscopy for most systems at $z > 2.8$ (and detections of \hb\ are often infeasible due to its faintness in low-luminosity galaxies), photometric techniques have been developed to measure \ha\ and other rest-frame-optical emission lines in longer wavelength imaging data.  For example, the existence of bright, high-equivalent width (EW) optical emission lines in $z\gtrsim4$ galaxies is inferred by measuring strong excesses in broad-band \textit{Spitzer}/IRAC Ch.1 (3.6 $\mu$m) and/or Ch. 2 (4.5 $\mu$m) photometry \cite[e.g.][]{2011ApJ...738...69S,2012ApJ...755..148G,2013ApJ...777L..19L,2014ApJ...784...58S,2016ApJ...823..143R,2016MNRAS.461.3886R}.  These studies have demonstrated that the typical rest-frame EWs of \oiii\ and \ha\ in $\sim$ $L^{\star}$ galaxies at high $z$ often exceed 300 \AA.  Moreover, the highest-EW sources at $z>7$ have ionizing photon production efficiencies that are a factor of 2.5 higher than the ``canonical'' value, implying a diversity in the ionization properties of the full galaxy population and that strong line emitters could be important contributors to cosmic Reionization \citep{2017MNRAS.472..772M,2017MNRAS.464..469S}.

The vast majority of the total galaxy population, namely those at sub-$L^{\star}$ UV luminosities, are much more difficult to understand at these redshifts.  The low spatial resolution (FWHM > 1.5 arcsecond) and broad filter widths ($>$ 6000 \AA) make shallow IRAC spectro-photometry for individual objects useful only when they are isolated and have relatively bright emission lines.  Photometric stacking, though, can be used in order to detect fainter emission lines so long as there are no nearby contaminating sources, or those sources can be accurately modeled.  As shown in \citet{2019AA...627A.164L}, this work can be extended to faint (\muv\ $>$ $-$18; 0.05 $L^{\star}$ at $z=4$) galaxies when using MUSE spectroscopy and ultra-deep \textit{Spitzer}/IRAC imaging with new techniques for modeling contamination from nearby sources \citep{2015ApJS..221...23L}.

Moving to even fainter sources is necessary to fully understand the ``budget'' of ionizing photons in the early Universe.  Sources fainter than \muv\ $\approx$ $-$17 do not appear in the deepest broad-band imaging taken with \textit{Hubble} unless they have been gravitationally lensed.  Yet, an abundant population of sources with \muv\ $\approx$ $-$15 (0.01  $L^{\star}$) at $z \approx 2.9-6.7$ have recently been discovered spectroscopically via \lya emission in un-targeted surveys with the MUSE spectrograph \citep{Bacon2017,2018ApJ...865L...1M}.  Their bright \lya emission and faint UV continuum (detected only in stacks) implies that they have extreme \lya EWs, in excess of 150 \AA.  Their number density is consistent with a simple extrapolation from the general population of star-forming \lyans-emitters at these redshifts.  Their high EWs can only be produced in stellar populations with ages less than 10 Myr and metallicities less than a few per cent Z$_{\odot}$ \citep[e.g.][cf. \citeauthor{1991ApJ...370L..85N} \citeyear{1991ApJ...370L..85N}]{2010AA...523A..64R,2017MNRAS.465.1543H}, physical conditions which are also conducive to efficiently producing ionizing photons.

Here we aim to study the ionizing photon production efficiency in these high-EW \lya emitters (LAEs) by using stacked \textit{Spitzer}/IRAC photometry to measure the \ha\ emission.  We can put these results into context by comparing to the numerous studies at similar redshifts that have probed more luminous galaxies: LAEs \citep[e.g.][]{2018ApJ...859...84H,2019AA...627A.164L}, \ha\ emitters \citep[HAEs;][]{2017MNRAS.465.3637M}, and continuum-dropouts \citep[e.g.][]{2016ApJ...831..176B}.  Our values will also be compared to larger samples of ``typical'' $\approx$ 0.3$-$3 $L^{\star}$ star forming galaxies at $z\approx2$ \citep[e.g.][]{2018ApJ...855...42S}, and theoretical stellar population models for the evolution of \xii\ with physical properties.

This article is organized as follows.  In Section \ref{sec:data} we describe the MUSE and IRAC datasets used for this study.  In Section \ref{sec:xi} we discuss how the data are used to determine the \ha\ luminosity and hence \xii, with a discussion of these results given in Section \ref{sec:discussion}.  Finally, in Section \ref{sec:conclusion} we summarize the primary results and give an outlook for future studies.  We adopt a flat $\Lambda$CDM cosmology ($\Omega_m=0.3$, $\Omega_\Lambda=0.7$, and H$_0=70 ~$km s$^{-1}$ Mpc$^{-1}$) and AB magnitudes \citep{1974ApJS...27...21O} throughout.

\section{Data}
\label{sec:data}
Using data from the MUSE spectrograph \citep{2010SPIE.7735E..08B} on the \textit{Very Large Telescope}, we select high-EW LAEs from the MUSE UDF survey \citep{Bacon2017,Inami2017}.  This survey covers approximately 9 square arcminutes to a depth of 10 to 30 hours at optical wavelengths (4750 $-$ 9300 \AA).  The un-targeted nature of MUSE spectroscopy means that objects do not need to be pre-selected based on their continuum properties, and we can search the data cubes for spectral features such as emission lines without requiring the detection of a continuum counterpart.  From these catalogs, we only consider LAEs in the range $3.829 < z < 4.955$ with confident redshifts (i.e. \texttt{CONFID} of 2 or greater), where \ha\ emission lies within the IRAC Ch. 1 bandpass.

In order to select a sample of high-EW LAEs, we must select not only on the \lya flux but also on the UV continuum level.  Using the aperture flux measurements from \citet{2018ApJ...865L...1M}, based on the combined 11-band \cite{XDF} and \cite{2018arXiv180601853O} reductions, we select only MUSE LAEs from the \citet{Inami2017} catalog that have no $>$3-$\sigma$ HST detections in bands redwards of \lyans.  To wit, objects detected only in the HST band that contains \lya are still included in the sample.  This results in a sample of 41 objects.  This selection differs from that presented in \citet{2019AA...627A.164L}, which is also based on \lyans-emitters from the MUSE UDF survey, as they require continuum detections for individual objects.  The threshold of 3-$\sigma$ compared with 5-$\sigma$ as in \citet{2018ApJ...865L...1M} is to restrict ourselves to only the highest-EW sources by limiting the acceptable detection limit in the UV continuum \cite[the average \textit{F850LP} 3-$\sigma$ magnitude limit in the field is approximately 30.0;][]{XDF}.  We note that the qualitative conclusions of this work are independent of this choice.

\begin{figure*}
\begin{center}
\includegraphics[width=.9\textwidth]{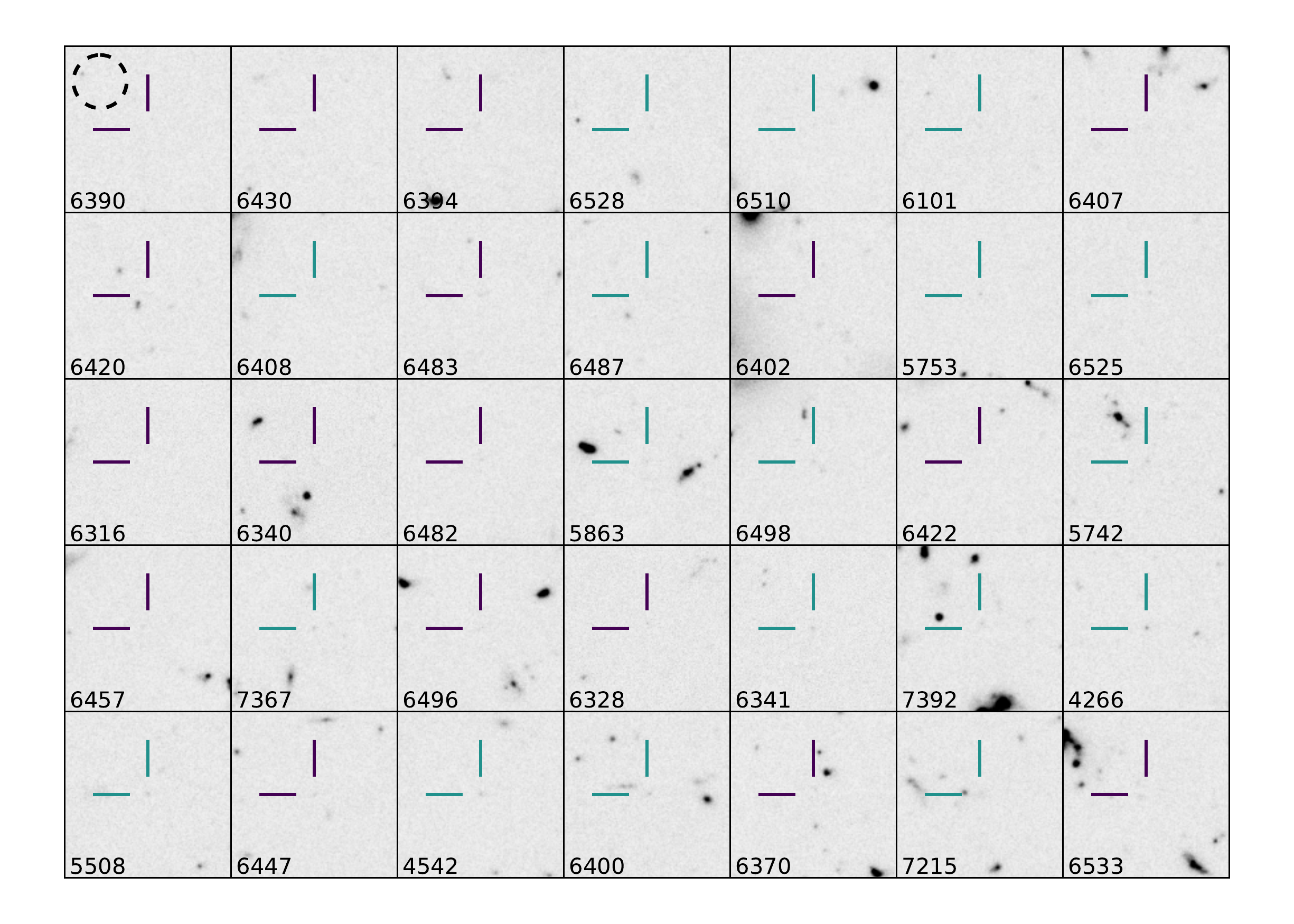} 
\end{center}
\caption{HST/ACS $F775W$ cutouts showing the $z\approx4-5$ LAEs used in this study.  Purple crosshairs denote objects that are not detected in any HST photometric band at a significance greater than 3-$\sigma$ \citep[cf.][]{2018ApJ...865L...1M}, while the blue crosshairs denote an object that is detected \textit{only} in the HST photometric band that contains \lya emission (i.e. $F775W$ and/or $F606W$).  Cutouts are 5 arcsec on a side, and the length of the arms of the crosshairs are 1 arcsec.  The dashed black circle in the upper-left panel shows the FWHM of the IRAC Ch. 1 PSF (1.6 arcsec diameter).}
\label{fig:cut}
\end{figure*}

A key addition to the HST photometry is near-infrared observations with \textit{Spitzer}/IRAC.  Here we utilize the combination of data from the ``GOODS Re-ionization Era wide-Area Treasury from \textit{Spitzer}'' (GREATS, PI: I. Labb\'{e}; M. Stefanon, in prep.) as well as all previous IRAC observations taken over the larger area around the UDF \citep{2015ApJS..221...23L,2019AA...627A.164L}.  The average exposure time of the IRAC Ch. 1 imaging is 167 hours \cite[26.6 magnitude using the photometric procedure of][5-$\sigma$]{2015ApJS..221...23L}, with some regions in the UDF receiving 278 hours of coverage.  In IRAC Ch. 2 the exposure time is similar (average 139 hours, deepest 264 hours, 26.5 magnitude).

From our candidate sample of 41, we manually remove 6 objects with severely contaminated IRAC photometry, and those sources with significant residual flux in the cleaned images (see Section \ref{sec:phot}).  This leaves a final sample of 35 high-EW LAEs with a median redshift of $z =$ 4.52, for which HST/ACS $F775W$ cutouts are shown in Figure \ref{fig:cut}.  This represents 19 per cent of all LAEs in this redshift range from the MUSE UDF catalogs \citep[][and Bacon et al. in prep.]{Inami2017}.  Based on the UV continuum slope measured for these galaxies described in Section \ref{sec:dust} and the median \lya flux from the PSF-weighted MUSE spectra \cite[which does not include extended emission;][]{Inami2017}, the average rest-frame \lya EW of the sample is 259 \AA\footnote{A 5-$\sigma$ continuum detection limit as in \citet{2018ApJ...865L...1M} would have resulted in a median rest-frame \lya EW of 162 \AA.}.

\subsection{De-blended IRAC photometry}
\label{sec:phot}

The large spatial point spread function (PSF) of IRAC (FWHM $\approx$ 2 arcsec) presents a challenge when dealing with compact sources in crowded fields.  This is especially true in deep imaging data, where the wings of the PSF from neighboring sources often overlap.  To correct for this effect, a higher spatial resolution image can be combined with a model PSF in order to de-blend the photometry in a crowded field.  In this case, we use HST $F850LP$ imaging as the high-resolution prior image and follow the procedure of \citet{2015ApJS..221...23L} and \citet{2019AA...627A.164L}. We create a model IRAC flux distribution for all continuum-detected sources in the field, which we then subtract from the IRAC data.  We are therefore left with a residual image that should only contain flux from the primary (continuum-undetected) source

All 12 $\times$ 12 arcsecond cutouts are visually inspected for defects, which can be caused by e.g. poorly-modeled bright sources or can occur in exceptionally crowded fields.  Any object that is deemed to have contaminated photometry due to model residuals is removed from the sample.  We note that these contamination effects are independent of the intrinsic properties of the primary source.  The resulting ``clean'' sample has residuals in the range of 0.01 to 0.3 nJy (cf. the mean from the ``contaminated'' sources of 0.6 nJy), determined via the standard deviation of pixel values within an annular region centred on the source and extending from 1.5 to 3 arcseconds.

\section{Determination of $\xi_{ion}$}
\label{sec:xi}

The three ingredients required to measure \xii\ are the flux of \ha\ (a proxy for the intrinsic ionizing photon production rate), the flux of the UV, and the escape fraction of ionizing photons:  
\begin{equation}
\label{eq:xi}
\xi_{\textrm{ion}} (\textrm{Hz erg}^{-1})= Q(H^0) / L_{\textrm{UV,int}}
\end{equation}
where $Q(H^0)$ is the intrinsic rate of ionizing photons with units of s$^{-1}$ and the intrinsic UV luminosity, $L_{\textrm{UV,int}}$, has units of erg s$^{-1}$ Hz$^{-1}$.  Below we outline the determinations of each of these quantities from our data.  

\subsection{\ha\ from IRAC stacking}
\label{sec:ha}

\begin{figure*}
\begin{center}
\includegraphics[width=.99\textwidth]{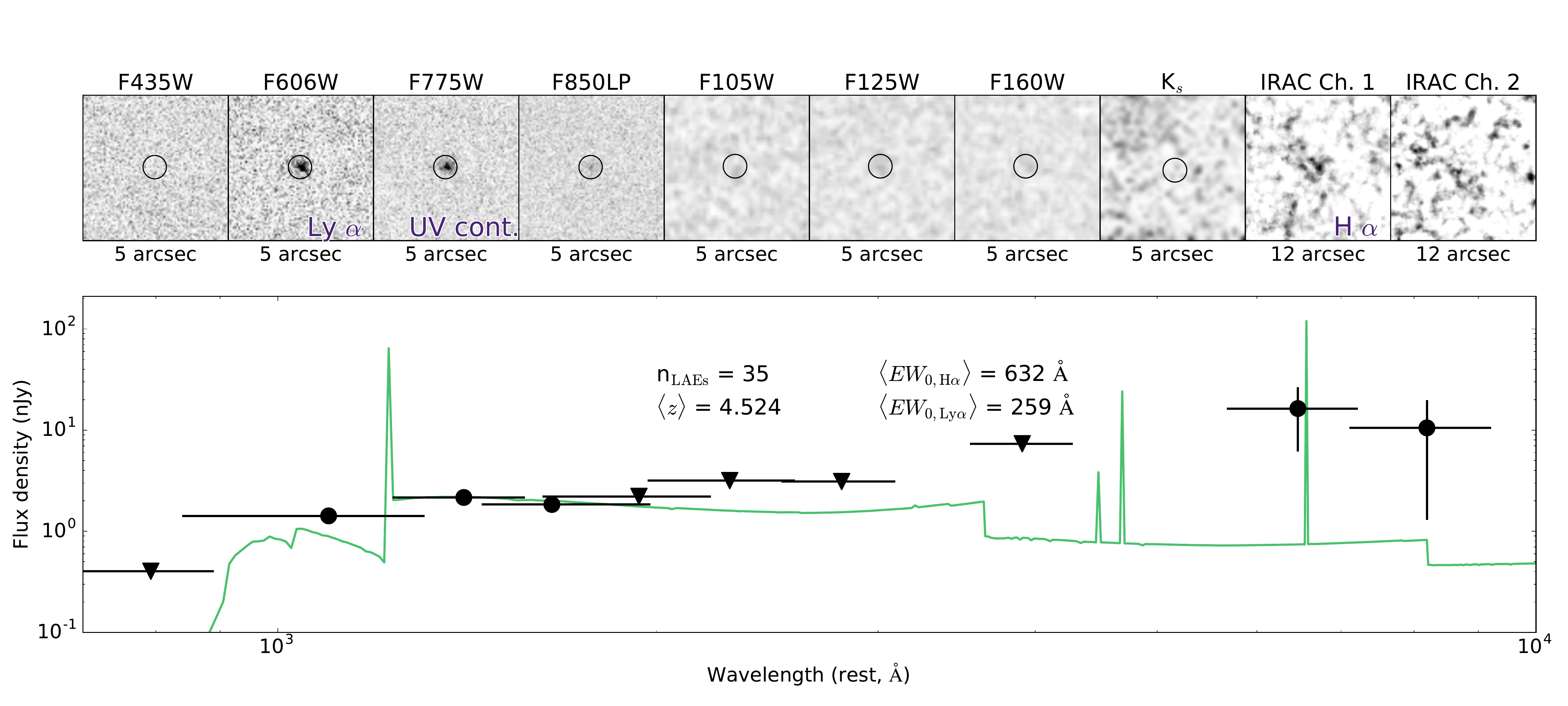} 
\caption{(Top) Mean-stacked HST, $K_s$, and IRAC photometry for the 35 high-EW LAEs in the sample.  The HST/$K_s$ cutouts are 5 arcseconds on a side, while the IRAC cutouts are 12 arcseconds.  The apertures shown are used for the photometric measurements of \muv\ from HST (0.4 arcsecond radius). (Bottom) Total observed restframe photometry and sample properties of the high-EW LAEs used in this study, as described in the text.  Upper limits (3-$\sigma$) are denoted with downward-facing triangles. For illustrative purposes, we also plot an arbitrarily-scaled theoretical model for a single stellar population 2 Myr after a burst of star formation, with a metallicity of 0.001 (see Section \ref{sec:discussion} for details).  Lines of Hydrogen and Helium are included in the model spectrum, as is the average intergalactic medium transmission at this redshift from \citet{2014MNRAS.442.1805I}.}
\label{fig:stack}
\end{center}
\end{figure*}

We create a three-dimensional array of the cleaned IRAC cutouts for the LAEs, each of which are centered on the peak of the \lya flux.  We calculate the mean of the array at each spatial pixel, incorporating a sigma-clipping procedure with a threshold of 2-$\sigma$ in order to ensure that individual bright pixels do not dominate the mean.  This ``stacking'' procedure results in a two-dimensional image, representing an average of the input sources.  As the IRAC PSF varies across the field due to the combined nature of the GREATS dataset \cite[see Section 3.2 of][]{2015ApJS..221...23L}, we also create a 3D array of the local PSFs at the position of each LAE.  This stack of PSFs is combined in the same way as the stack of the science data arrays, using the same mask derived from sigma-clipping.  We fit this PSF to the stacked image to determine the total flux, under the assumption that our sources are unresolved compared to the 2 arcsecond IRAC PSF (cf. the HST images).  The fit includes uncertainties in the flux determined from stacking the IRAC noise images in the same way as the data.  This stacking procedure is used to measure the photometry and create the images shown in Figure \ref{fig:stack}.  

For the remainder of this work, we deal with the distribution of the output IRAC Ch. 1 and Ch. 2 fluxes using random subsets of the 35 individual LAEs.  Namely, we create a series of 10,000 data arrays made up of a random set of 35 LAEs, where we sample the full set with replacement.  Each of these arrays is analyzed as above to determine flux in Ch. 1 and Ch. 2.   This bootstrap procedure allows us to measure the distribution of \ha\ fluxes within the sample while taking into account photometric uncertainties. 

We determine the \ha\ fluxes for each of the 10,000 bootstrap iterations by subtracting the Ch. 2 flux from the Ch. 1 flux, assuming Ch. 2 probes the underlying stellar continuum (see Section \ref{sec:cont}).  We attribute all of the excess flux in Ch.1 to H$\alpha$ emission: in the models of \citet{2016MNRAS.462.1757G}, for metallicities $\lesssim$ 10 per cent Z$_{\odot}$ the ratio of \nii\ (\sii) to \ha\ never exceeds 0.01 (0.1) over a wide range in physical ISM conditions.  While a detailed determination of the metallicities of these systems is beyond the scope of this paper and will require additional spectroscopic data, we stress that only models of normal stellar populations with extremely young ages ($<$ 3 Myr) and/or low metallicities ($Z < 0.008$; 0.4 Z$_{\odot}$) can reproduce such large \lya EWs \cite[][and Section \ref{sec:discussion}]{2017MNRAS.465.1543H,Hashimoto2017}.  Similar conclusions are drawn by \citet{2016ApJ...832..171T} for the faintest $z\approx2.5$ LAEs, where they have spectroscopic access to strong optical emission line ratios.

\subsubsection{The continuum level around \ha}
\label{sec:cont}
In order to properly measure the emission line flux from a photometric broad-band measurement, we need to establish the local spectral continuum level.  For the aforementioned studies that perform measurements of \ha\ EW at high-$z$, the continuum level at the position of \ha\ is typically established with photometric detections in the bands bluewards/redwards of \ha, such as the $K_s$-band or IRAC Ch. 2.  In our case, the ground-based $K_s$-band data \cite[from zFOURGE;][]{2016ApJ...830...51S} are not deep enough for reliable detections of the continuum flux bluewards of \ha.  As the IRAC Ch. 2 stacks are deeper and are subject to similar systematics as the Ch. 1 data, we use them to constrain the continuum redwards of \ha\ directly.

We assume a flat continuum slope in $f_{\nu}$ \citep{2011ApJ...742..111V}.  Any residual contamination, if present in both Ch. 1 and Ch. 2, would not bias our determination of the line flux (and hence \xii) since this is determined via the difference between both measurements.  Furthermore, residual contamination would lower the measured \ha\ EW for a spectrum that is flat in $f_{\nu}$.  We also assume no contribution to the Ch. 2 flux from emission lines such as \siii, which is present for the $z < 4.51$ subset of LAEs.  As shown in \citet{2019AA...627A.164L}, \siii\ $\lambda$9069 can be strong in LAEs, with EW values in excess of 100 \AA.  Constraints on \siii\ do not currently exist for galaxies as faint as the LAEs probed here, but at metallicities below 0.3 \zsol\ the ratio of \siii\ to \ha\ is predicted to be less than 0.05 \citep{2001MNRAS.323..887C}.  Furthermore, any contribution of \siii\ to the IRAC Ch. 2 photometry would mean that we are over-estimating the continuum level and hence under-estimating the strength of \ha\ (and \xii).

We determine a mean \ha\ EW for the sample of 632 \AA.  The 68 per cent confidence interval, weighted by the signal-to-noise of each bootstrap measurement of the \ha\ EW, is between 210 and 1600 \AA.  The large range in EW measurements is primarily driven by the lower signal-to-noise in the IRAC Ch. 2 stacks, which is typically a factor of 1.9 lower than that derived in IRAC Ch. 1.  In the case of measurements at low-EW (i.e. $<$ 210 \AA), the typical signal-to-noise in the EW determination is 2.1.

The mean measured star formation rate (SFR) from the \ha\ luminosity is 1.2 M$_{\odot}$ yr$^{-1}$ (0.4 $-$ 2 M$_{\odot}$ yr$^{-1}$; 68 per cent) when using the \citet{2011ApJ...737...67M} conversion.  This is larger than the implied UV-based SFR of 0.1 M$_{\odot}$ yr$^{-1}$, likely due to the fact that \ha\ emission probes star formation on shorter timescales than the UV, namely $<$ 10 Myr, which is important given the implied young ages for these systems: see Section \ref{sec:discussion}.

\subsection{UV luminosity from HST stacking}

Following the same procedure as for the IRAC stacks, for each bootstrap sample we create stacks of the HST images in order to determine \muv; for more details on this procedure, see \citet{2018ApJ...865L...1M}.  For the majority of galaxies in our sample, ACS/$F775W$ probes the rest-UV continuum directly.  However, for the highest-$z$ objects this band contains flux from \lyans: 3 of the 35 objects have \lya at a position where the throughput of $F775W$ is greater than 33 per cent. The major results of this paper are consistent within the errors ($<$ 1-$\sigma$) when these higher-$z$ sources are included.  Throughout, we assume that $F775W$ probes the UV continuum alone for our sample.  Any potential contamination to this flux from \lya emission would mean we are underestimating \xii.  Additionally, an extrapolation assuming a UV continuum slope of -2.5 from the measured $F850LP$ magnitude is consistent with the measured $F775W$ flux (see Section \ref{sec:dust}).  Such a continuum slope is expected for faint LAEs \citep{2017MNRAS.465.1543H}, and therefore this implies $F775W$ is a good tracer of the rest-frame UV continuum.  We convert the measured \muv\ into a luminosity assuming the median redshift for the LAEs in the bootstrap iteration.

\subsection{Ionizing photon escape fraction}
\label{sec:fesc}
The escape fraction of ionizing photons relates the observed ionizing photon flux with the intrinsically-produced ionizing photon flux:

\begin{equation}
f_{\mathrm{esc}}^{\mathrm{ion}} = \frac{Q_{\mathrm{obs}}(H^0)}{Q(H^0)}.
\end{equation}
For brevity, we will refer to \fion\ as \fesc\ throughout the remainder of the text.  This parameter is difficult to measure directly as $Q(H^0)$ is not a readily-observable quantity.  Studies such as \citet{2010ApJ...724.1524O} suggest using SED fitting to the broadband photometry to estimate \fesc\ indirectly.  At $z=6-7$ they find \fesc\ to be consistent with zero and constrain it to be $<$ 0.6.  Using a similar method, \citet{2018ApJ...859...84H} fit a relation to \fesc\ versus EW$_{\mathrm{Ly\alpha}}$, finding a value consistent with zero for rest-frame EWs in excess of 100 \AA.  We therefore assume \fesc\ = 0 throughout, but note that a higher value of \fesc\ will result in a larger \xii.

\subsection{Dust attenuation}
\label{sec:dust}
The true amount of dust extinction in these systems is difficult to establish with SED fitting as even in stacked images we only observe the rest-UV portion of the galaxies in addition to the \ha\ emission.  Many studies, particularly at these redshifts $z\approx4-5$, instead rely on a measurement of the UV continuum slope $\beta$ and assume a dust law, as in \citet{1999ApJ...521...64M}. We measure a median $\beta$ = -2.43 from power-law fits to the rest-UV photometry from each of the bootstrap iterations, with a 68\% confidence interval from -3.00 to -2.04, from a power-law fit to the stacked HST photometry for the full stacked sample.  We defer a detailed treatment of the $\beta$ slopes of the full sample of high-EW LAEs to a forthcoming paper, but we would like to highlight that the intrinsic stellar UV continuum slope for systems with the highest \lya EWs must be even bluer than the observed value due to the contribution of nebular continuum emission \citep{2010AA...523A..64R}. 

For $\beta <$ -2.23, \citet{1999ApJ...521...64M} estimate zero dust correction.  As our best-estimate of $\beta$ (and 68 per cent of all bootstrap iterations) is below this, we determine that we do not need to include an additional term for dust attenuation when measuring line fluxes.  Although our measurement is uncertain due to the intrinsic faintness of the sample, we \textit{a priori} expect such a correction to be negligible based on results from (UV-brighter) LAEs at similar or higher redshifts \cite[e.g.][]{2015MNRAS.454.1393S,2017MNRAS.465.1543H,2018ApJ...859...84H} and the result from \citet{2016ApJ...832..171T} which shows an anti-correlation between nebular reddening and EW$_{\mathrm{Ly\alpha}}$. Similarly, \citet{2018arXiv180909637T} measure the dust attenuation via the Balmer decrement for high-EW \oiii- and \ha-emitters, finding negligible extinction for the highest-EW objects.  Finally, such a result is also expected when considering the implied low stellar masses and gas-phase metallicities required to power the \lya emission \citep{2010MNRAS.409..421G}.

\subsection{The sample distribution of \xii}
\label{sec:xidist}

For each bootstrap iteration, we convert the \ha\ flux into a luminosity assuming the median redshift for the LAEs that contributed to the stack.  We then convert the luminosity (without a dust correction, as explained above) into the ionizing photon production rate $Q(H^0)$ assuming Case B recombination at a temperature of 10$^4$ K according to:

\begin{equation}
\left[\frac{Q(H^0)}{\textrm{s}^{-1}}\right] \times (1-f_{\textrm{esc}}) = \left(\frac{L_{H \alpha}}{\textrm{erg s}^{-1}}\right) \times 7.37 \times 10^{11},
\end{equation}
as in e.g. \citet{2011ApJ...737...67M}.

We can therefore calculate \xii\ according to the usual formula given in Equation \ref{eq:xi}.  The weighted mean of the distribution from the bootstrap iterations assuming zero \fesc\ is log$_{10}$ (\xii\ / Hz erg$^{-1}$) $=$ 26.28, with a 68 per cent confidence interval from 25.89 $-$ 26.56, weighted by the individual uncertainties on each measurement.

\begin{figure*}
\begin{center}
\includegraphics[width=.95\textwidth]{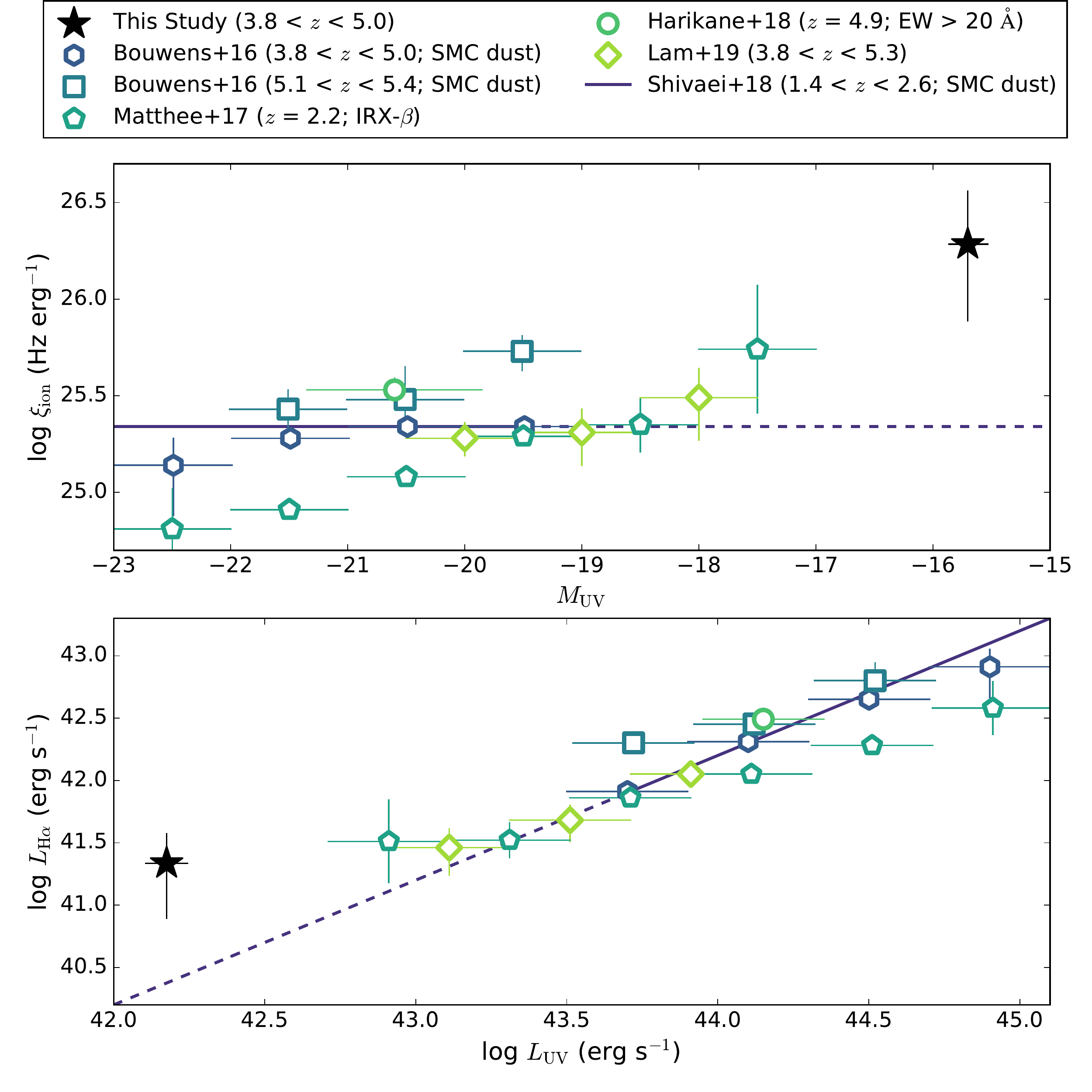}
\caption{\xii\ versus \muv\ for this sample, emission line-selected literature samples of \citet{2017MNRAS.465.3637M,2018ApJ...859...84H,2019AA...627A.164L}, and the continuum-selected sample of \citet{2016ApJ...831..176B}. In the case of \citet{2018ApJ...855...42S}, the dashed line denotes an extrapolation below their faintest bin.  The legend describes which result we use from the literature when the authors present multiple values for \xii\ depending on e.g. their assumed dust law.  The error bars on our data points show the (marginalized) sample distribution based on bootstrapping (see Section \ref{sec:ha}), whereas the error bars on \muv\ from the literature sample reflect the bin width used in each study.  The bottom panel isolates the two individual components of \xii, as the axes in the top panel are strongly correlated: the quoted uncertainties in \xii\ for the literature samples do not typically include the width of the \muv\ bins.  Our high-EW LAEs are have higher ionizing photon production efficiency for their continuum (UV) luminosity than the literature samples.}
\label{fig:xi_muv}
\end{center}
\end{figure*}

In Figure \ref{fig:xi_muv}, we show our estimate of \xii\ versus \muv, compared to literature samples of line-selected samples \cite[HAEs and LAEs][]{2017MNRAS.465.3637M,2018ApJ...859...84H,2019AA...627A.164L} and continuum-selected galaxies \citep{2016ApJ...831..176B} at similar redshifts, as well as ``normal'' $M_{\star} > 10^9$ M$_{\odot}$, $L \approx$ 0.3$-$3 $L^{\star}$ star-forming galaxies at $z\approx2$ from \citet{2018ApJ...855...42S}.  We use their value derived from an SMC extinction curve for consistency with the other studies presented here.  The bottom panel shows the relationship between the measured luminosities of \ha\ and the rest-frame-UV, which are the two components that go into \xii; any trend in \xii\ versus \muv\ could simply be because the UV luminosity goes into both the ordinate and the abcissa.  While the information content in both panels is the same, the axes in the top panel are necessarily correlated.

While all of the LAE- and LBG-derived data points are within 2-$\sigma$ (assuming the bin widths correspond to a measurement uncertainty on $L_{\mathrm{UV}}$) of the \citet{2018ApJ...855...42S} relation or its extrapolation in either $L_{\mathrm{H \alpha}}$ or $L_{\mathrm{UV}}$, the point derived here is more significantly above the relation for $L_{\mathrm{H \alpha}}$ and  $L_{\mathrm{UV}}$.  Therefore, relative to their UV luminosity, the MUSE high-EW LAEs are more efficient at producing ionizing photons than the general population of more luminous LBGs and LAEs at these redshifts.

\subsection{The \lya escape fraction}
\label{sec:lyafesc}
As discussed in Section \ref{sec:fesc}, it is difficult to directly measure \fesc\ in these galaxies, although based on other samples we expect it to be low.  We can, however, measure the escape fraction of \lya photons:

\begin{equation}
f_{\mathrm{esc}}^{\mathrm{Ly\alpha}} = \frac{L_{\mathrm{Ly\alpha}}^{\mathrm{obs}}}{L_{\mathrm{Ly\alpha}}^{\mathrm{int}}} = \frac{L_{\mathrm{Ly\alpha}}^{\mathrm{obs}}}{8.7 \times L_{\mathrm{H\alpha}}^{\mathrm{int}}}
\end{equation}
where the superscripts \textit{obs} and \textit{int} refer to the observed and intrinsic luminosities, respectively.  This equation is valid for Case B recombination with a temperature of 10$^4$ K \cite[see ][]{2015ApJ...809...19H, 2015ApJ...809...89T}.  Based on our observations, we derive a mean \flya value of  0.217, with a 68 per cent confidence interval of 0.067 $-$  0.333 based on our bootstrap iterations.

\section{Discussion}
\label{sec:discussion}

Qualitatively, several studies have found a trend towards having higher \xii\ values in galaxies with bluer UV continua, a proxy for galaxies with more dominant young stellar populations \citep{2015MNRAS.451.2030D,2015ApJ...811..140B,2016ApJ...831..176B}.  However, \citet{2017MNRAS.465.3637M} show that this is a product of using the UV slope itself (or galaxy SEDs that predominantly trace the rest-UV) as a proxy for dust attenuation.    This is also sensitive to the assumed dust model, as can be seen in Figure 4 of \citet{2018ApJ...855...42S} where using different dust laws can produce a marginally positive or a negative correlation between \xii\ and \muv.  Nevertheless, \citet{2017MNRAS.465.3637M} find that \xii\ increases with decreasing UV luminosity, increasing specific star formation rate, and increasing \ha\ EW in their sample of $z\approx2$ \ha\ emitters.

\begin{figure}
\begin{center}
\includegraphics[width=.45\textwidth]{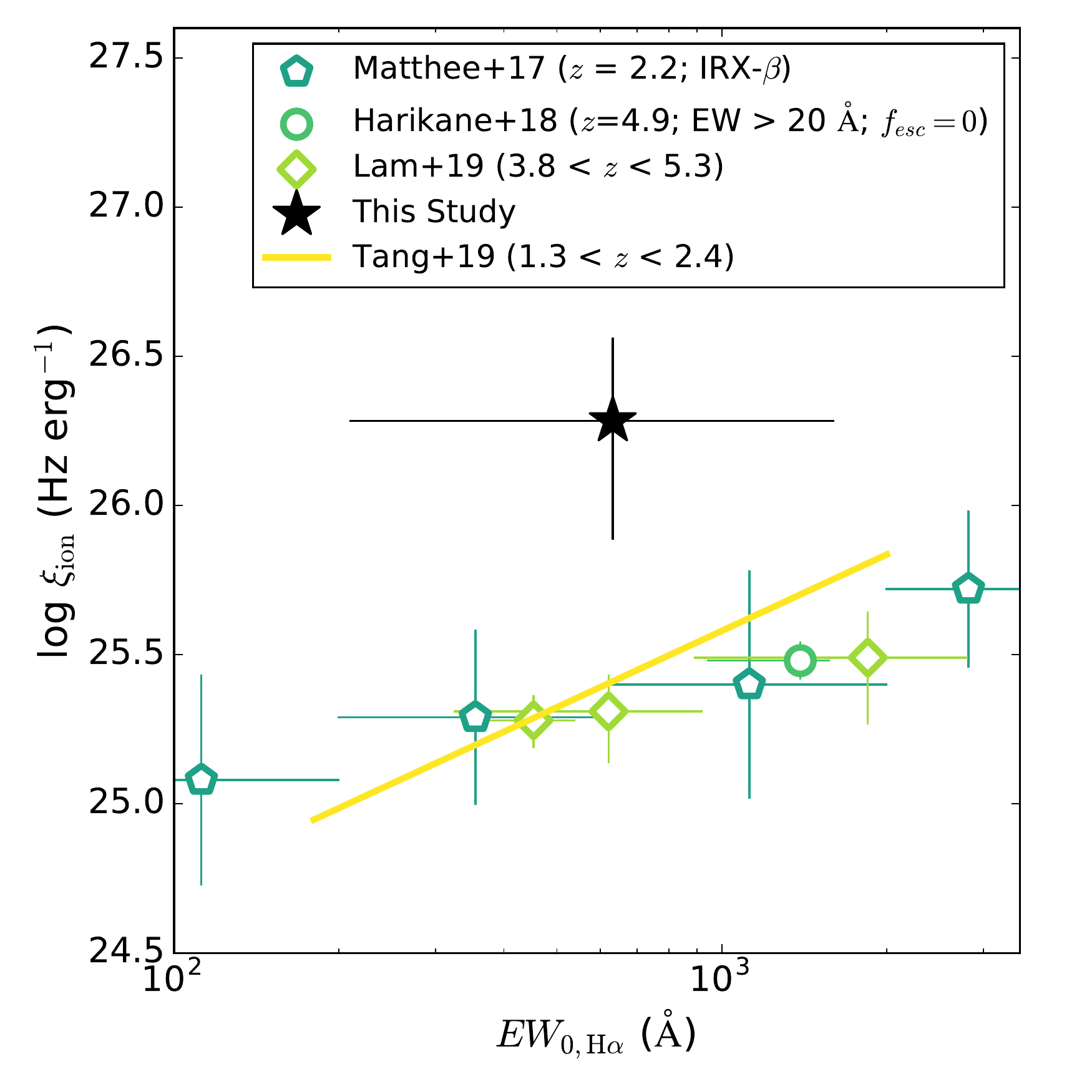} 
\caption{Restframe \ha\ EW versus \xii\ for this sample, compared to the literature observations of other line emitters (LAEs and HAEs).  The error bars are not independent as a higher \ha\ EW necessarily implies a higher \xii.  The relationship between the EW of \ha\ and \xii\ is consistent between the literature samples, with a strong trend to higher \xii\ at higher \ha\ EWs.  As the \ha\ EW is inversely proportional to the age, this smooth trend suggests that younger ages are the main driver of elevated \xii.  Our elevated value at fixed \ha\ EW suggests that these LAEs have a lower (gas-phase) metallicity than other samples presented in the literature (see Figure \ref{fig:raiter}).}
\label{fig:xi_ew}
\end{center}
\end{figure}

In Figure \ref{fig:xi_ew} we compare the equivalent width of \ha\ with the derived \xii\ value for the emission line-selected samples.  In addition, we plot the fit to the relation from \citet{2018arXiv180909637T} derived from $z\approx1-2$ ``extreme emission line galaxies'' (EELGs), which are selected purely on the basis of high-EW optical emission lines (\oiii\ and/or \ha).  These systems have gas-phase metallicities $<$ 0.3 Z$_{\odot}$ and mass doubling times on the order of 100 Myr \citep{Maseda13,maseda14}.  While these physical properties are broadly similar to those of literature samples of LAEs at higher redshifts, the offset compared with the high-EW LAEs presented here argues for a difference in \xii\ at a fixed age. This is due to the fact that the EW of Balmer lines scales inversely with the age of the stellar population in these types of systems \citep[e.g.][]{2011ApJ...742..111V}.  Here and throughout, we specifically refer to the light-weighted age, which is dominated by the most recent generation of star formation: an older ($>$ 1 Gyr) stellar population could contribute to the total mass of the galaxy but would not contribute significantly to the UV or \ha\ luminosities.  The highest-EW objects in the \citet{2018arXiv180909637T} sample, with ages $<$ 10 Myr, have the highest values of \xii.  As these galaxies and the high-EW LAEs presented here have little to no dust attenuation (see Section \ref{sec:dust}), any correlation between \xii\ and age is not driven by using the rest-UV to correct for dust attenuation.  At a fixed \ha\ EW, our sample presents a larger value of \xii\ than the literature LAE/HAE results, implying a lower gas-phase metallicity.  However, in all samples a trend with higher \xii\ at younger ages is found.

\begin{figure*}
\begin{center}
\includegraphics[width=.95\textwidth]{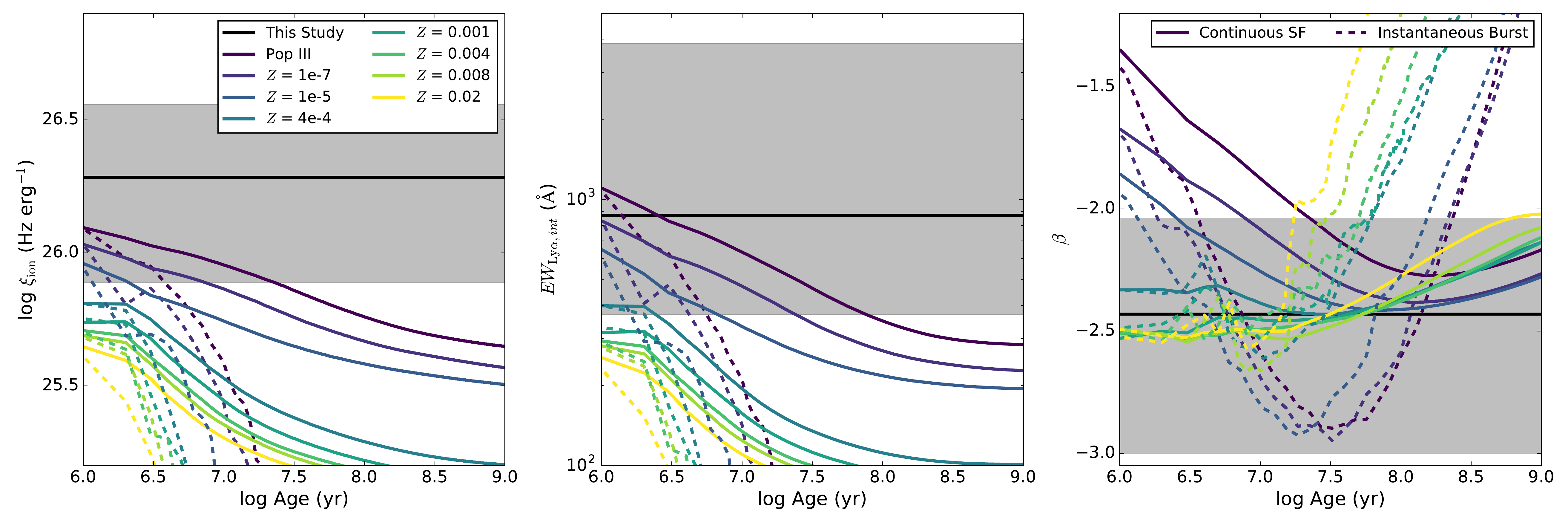} 
\caption{Predicted \xii\ (left), intrinsic \lya EW (i.e. corrected for the \lya escape fraction; centre), and UV continuum slope ($\beta$; right) from the \citet{2010AA...523A..64R} stellar population models with constant star formation (solid lines) or an instantaneous burst (dotted lines) at different metallicities (colors), compared to the results presented here.  The shaded regions denote our 68 per cent confidence interval in \xii\ (left), the intrinsic \lya EW for the sample (centre; including the uncertainty on the \lya escape fraction), and $\beta$ (right), all from the bootstrap iterations described in Section \ref{sec:ha}}.  We quantify the goodness-of-fit for each combination of star formation history, age, and metallicity in Figure \ref{fig:raiterfit}.
\label{fig:raiter}
\end{center}
\end{figure*}

\begin{figure*}
\begin{center}
\includegraphics[width=.95\textwidth]{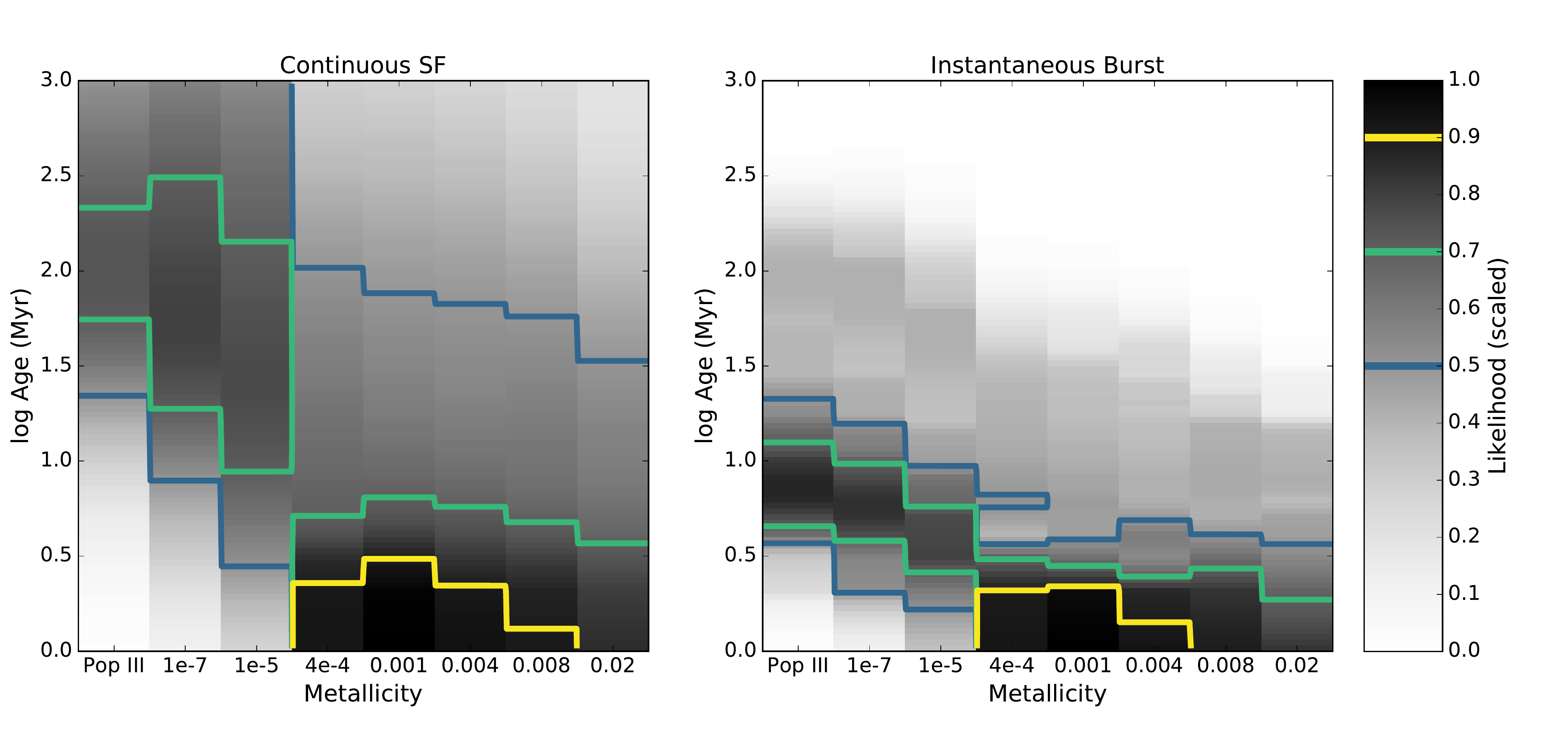} 
\caption{Scaled likelihood of each of the \citet{2010AA...523A..64R} models to reproduce our observations as a function of age and metallicity for a constant star formation history (left) and an instantaneous burst (right).  As each combination of metallicity, age, and star formation history produces an estimate for each of our three measured quantities (\xii, \lya EW, and $\beta$; Figure \ref{fig:raiter}), we can calculate a likelihood for that combination by comparing the predictions to our posterior probability distributions for each quantity.  The total likelihood is then the product of the three individual probabilities, scaled to a maximum of 1.  As shown by the colored contours, the most likely models have metallicities of 4$\times$10$^{-4}~-$ 0.008 and ages less than 3 Myr, regardless of the star formation history.  We cannot, however, conclusively rule out more metal-poor stellar populations with older ages ($>$ 10 Myr for constant star formation or $>$ 3 Myr for an instantaneous burst).}
\label{fig:raiterfit}
\end{center}
\end{figure*}

Qualitatively, for a constant star formation history \xii\ is roughly constant for the first 1 Myr before monotonically declining until the population is approximately 1 Gyr old, when it plateaus again (Figure \ref{fig:raiter}).  This is related to differences in the star-formation timescales probed by \ha\ and the UV, and the effect is strongest between 1 and 10 Myr \cite[or longer when binary stellar evolution is included in the models, e.g.][]{2016MNRAS.456..485S}.  For an instantaneous burst of star formation, the value of \xii\ drops strongly after 1 Myr.  At a fixed age (and hence UV luminosity) for both star formation histories, decreasing metallicity results in an increasing \xii\ or \ha\ luminosity.  So, while young ages result in higher \xii\ values for the (bursty) HAEs compared to the general population of star-forming galaxies, the observed offset in \xii\ at fixed \ha\ EW (and hence fixed age) for the high-EW LAEs shown in Figure \ref{fig:xi_ew} suggests that the metallicities must be lower and/or \fion\ must be higher.  Independent constraints on the metallicity from e.g. rest-optical spectroscopy will be required to determine the contribution of each.

In Figure \ref{fig:raiter}, we show the predictions from the \citet{2010AA...523A..64R} stellar population models for the evolution of the \ha, \lyans, and UV luminosities at different metallicities for a constant SFR and an instantaneous burst (their ``cs5'' and ``is5'' models, respectively).  These models use the stellar tracks, atmospheres, and prescriptions for nebular (line and continuum) emission from \citet{2003AA...397..527S}.  For the nebular emission, they assume ionization-bounded nebulae with constant electron temperature and densities, and that all photons are absorbed inside the \hii\ regions (i.e. \fion\ $=$ 0).  We consider models with a \citet{1955ApJ...121..161S} initial mass function, with a high-mass cutoff of 500 M$_{\odot}$.  

At each age and for a given star formation history, the fixed-metallicity models predict our observable quantities: \xii, the EW of \lyans, and the UV continuum slope $\beta$.  We can compare these predicted values to the posterior distributions derived from our bootstrapping procedure and derive a (relative) likelihood for each model.  This grid of likelihoods is shown in Figure \ref{fig:raiterfit}.  While we do not have independent constraints on the star formation history, in both cases the most likely models have metallicities of 0.001 (4$\times$10$^{-4}~-$ 0.008) and ages less than 3 Myr.  Lower metallicity models are also permitted with older ages, up to 300 Myr.  Although the specific metallicity threshold is dependent on our choice of stellar population model and IMF \cite[cf.][]{2016MNRAS.456..485S}, our median value of \xii\ is difficult to reproduce with any current set of models even when the high-mass cutoff of the IMF is set to 500 M$_{\odot}$ (T. Nanayakkara in prep.).  This could be related to the lack of very hard ionizing photons in these models which struggle to reproduce observations of high-$z$, low-metallicity galaxies \cite[e.g.][]{2019AA...624A..89N,2019ARAA..57..511K}.

\citet{2016ApJ...832..171T} determine a maximum \xii\ value of $10^{25.6}$ Hz erg$^{-1}$ at 7 per cent Z$_{\odot}$ from their sample of $z\approx2.5$ LAEs, some of which have \lya EWs in excess of 100 \AA\ and have UV magnitudes fainter than $-$18. Based on their measurement of a high ionization parameter across their full sample, they conclude that age alone does not drive the elevated \xii\ that they measure and hence these galaxies should be producing ionizing photons in a steady state.  Indeed, even at 100 Myr the models shown in Figure \ref{fig:raiter} with metallicities below 3 per cent Z$_{\odot}$ are elevated with respect to the canonical value.  Metallicity can indeed play a role in having an elevated \xii, as the extreme \lya EWs require metal-poor ($<$ 0.02 Z$_{\odot}$) stellar populations with ages of 10 Myr or less \cite[Figure \ref{fig:raiter} and][]{2010ApJ...724.1524O,2017MNRAS.465.1543H}.

The EW of \lya\ is observed to scale with \flya\ \cite[which is correlated with \fion, albeit with large scatter][]{2016ApJ...828...71D}, and for our sample of high-EW LAEs we would expect to have \flya\ values close to 1 \citep{2019AA...623A.157S}.  However, we do not observe such large values of \flya: as shown in Section \ref{sec:lyafesc}, the mean value is 0.217 assuming Case B recombination \cite[cf.][]{2010AA...523A..64R}. This matches the value of 0.3 found in \citet{2015ApJ...809...89T} despite that sample having a mean \lya\ EW of only 43 \AA. Interestingly, recent observations of an EELG at $z\approx1.8$ by \citet{2019arXiv190711733E} find \flya to be 0.097, which is also low for its \lya EW.  \citet{2019arXiv190809763J} do not find a strong correlation between \ha\ EW, \lya EW, and \flya in ``green pea galaxies,'' which have extreme ionization fields that are plausibly similar to the galaxies presented here.  They note that the highest \ha\ EWs reflect high intrinsic \lya production and hence a large escape fraction is not required to produce the observed EWs of \lyans.  In fact, the \citet{2019AA...623A.157S} model predicts the relationship between \flya and the EW of \lya to flatten with increasing \xii, such that we could expect a similar value of \flya to the one we measure.  The extreme galaxies we select here, then, do not require high values of \flya\ to have the very high EW values measured here (259 \AA; median).  In addition, \citet{2019MNRAS.484...39S} find that the escape of \lya photons lags behind the star formation activity by several tens of Myrs, hence our selection of the youngest star formation episodes via high-EW \lya could preferentially select periods of relatively low escape.  

\subsection{Alternatives to Star Formation}

So far we have only considered star formation as the source of ionizing photons in these systems.  Overall, with the current data it is difficult to determine what is the main source of ionizing photons inside these galaxies.  As previously mentioned, existing models of normal stellar populations can generally reproduce our observables (i.e. Figure \ref{fig:raiter}). Similar conclusions are drawn
by \citet{2016ApJ...832..171T} for $z \approx$ 3 LAEs, where they use strong optical emission line ratios to demonstrate that star-formation is
the dominant source of ionizing photons. In addition, we have demonstrated in \citet{2018ApJ...865L...1M} that the undetected sample plausibly
represent the high-EW extension of the distribution of LAEs at these redshifts, a population which is not dominated by AGN.

Thus, we do not believe that AGN are the only or even the dominant contributor to these systems. However, we cannot conclusively rule out some contribution, either as the primary source in a subset of objects or a secondary source in all objects. Further spectroscopy is required in both the restframe-UV and restframe-optical.

\section{Outlook and Conclusions}
\label{sec:conclusion}

In this work, we have identified a population of 35 $z\approx4-5$ high-EW LAEs using deep MUSE spectroscopic data in the UDF, with a median \lya EW of 249 \AA.  By combining this with ultra-deep \textit{Spitzer}/IRAC photometry, we have measured \ha\ emission with an EW of  632 \AA\ (210 $-$ 1600 \AA; 68 per cent) from these systems, and used it to calculate the ionizing photon production efficiency, \xii\ (Equation \ref{eq:xi}).  Our primary conclusions are as follows:

\begin{itemize}
\item{Using \textit{Spitzer}/IRAC photometry to $\approx$ 200 hour depth and sophisticated techniques for source deblending, we detect \ha\ emission in stacked data for objects with intrinsic UV magnitudes fainter than $-$16 (Figure \ref{fig:stack}).}
\item{The  mean value of \xii\ for these high-EW LAEs is $10^{26.28}$ Hz erg$^{-1}$ for an escape fraction of zero, with a 68 per cent confidence interval from $10^{25.89} - 10^{26.56}$ Hz erg$^{-1}$ (Section \ref{sec:xidist}).  At a UV magnitude of $-$15.7, this value is a factor of 8.7 in excess of the ``canonical'' value from literature studies of emission line-selected and continuum-selected samples at similar and higher redshifts, and also higher than the most extreme values for $z\approx7$ galaxies from \citet{2017MNRAS.464..469S} or local galaxies from \citet{2018MNRAS.479.3264C}.}
\item{While the values of \xii\ from the literature are consistent with a constant value at all \muv\ \cite[e.g.][]{2019AA...627A.164L}, our observation lies significantly above this relation (Figure \ref{fig:xi_muv}).  This naturally follows from selecting objects with bright \lya luminosities (and hence ionizing photon production rate) compared to the UV continuum, i.e. selecting on the EW of \lyans.}
\item{Based on the observed trend between \ha\ EW and \xii, as well as models of the evolution of \xii\ from stellar population synthesis, we determine that an elevated \ha\ luminosity compared with the UV luminosity is likely a natural consequence of a gas-phase metallicity between 4$\times$10$^{-4}$ and 0.008 and an age younger than 3 Myr, regardless of the star formation history (Figures \ref{fig:raiter} and \ref{fig:raiterfit}).}
\end{itemize}

Typical galaxies selected via the Lyman-break technique \citep[e.g.][]{2017ApJ...843..129B} require detections in the rest-UV, and hence are biased against the youngest ($<$ 10 Myr) galaxies which necessarily have not produced much UV flux.  This is true for LAE selections as well \citep[e.g.][]{2019AA...627A.164L}, even in samples derived from narrow-band imaging \citep[e.g.][]{2018ApJ...859...84H}, as the relative depth of the emission line detection threshold with respect to the continuum detection threshold prevents these studies from finding the youngest systems.  Our MUSE \lya selection, on the other hand, is closer to an \ha\ selection considering it is also a measurement of (emergent) ionizing photons; by requiring non-detections in the rest-frame-UV, we preferentially select the youngest systems.  With a selection based on maximal line emission and minimal continuum, it is logical that we have found a population of galaxies with elevated values of \xii\ compared with older, more massive galaxies.

We expect that observing continuum-faint LAEs with lower line EWs would result in a lower determination of \xii\ as these systems could have older stellar populations, higher metallicities, or both.  Indeed, lower-$z$ observations of \xii\ show significant scatter at low luminosities and stellar masses (M. Paalvast et al. submitted), which is potentially due to a distribution in the ages, star-formation histories, and metallicities in this regime.  These galaxies are currently unobservable in emission at high-$z$, but it is possible to measure the product of \fesc\ and \xii\ for systems detected in absorption down to \muv\ $\approx$ $-$16 \citep{2019MNRAS.483...19M} which could select a population of galaxies that are distinct from LAEs.

Various hydrodynamical simulations predict that the star formation histories of low-mass galaxies in the early Universe is episodic in nature, with periods of star formation followed by more quiescent phases \cite[e.g.][]{2014ApJ...792...99S,2015MNRAS.454.2691M}.  Although the precise duty cycle of these star formation episodes is unknown and could vary with properties such as the galaxy halo mass \citep{2011ApJ...742..111V}, a single galaxy could undergo multiple episodes of efficient ionizing photon production  without significantly building up its stellar mass \citep[e.g.][]{2015MNRAS.451..839D}.   Although a phase with extreme \lya EW may not occur during the lifetime of every galaxy, depending on their star formation and metal enrichment histories, all galaxies should produce an excess of ionizing photons at young ages for several Myrs. Further work is required to fully characterize the duty cycle of intense star-formation episodes at high-$z$, but the steep faint-end slope of the UV luminosity function and the observed high number density of high-EW LAEs implies that a significant number of such events should be taking place across cosmic time: at $z=4-5$, high-EW LAEs as selected here represent 6.5 per cent of the full galaxy population at a UV magnitude of $-$16, based on the \citet{2015ApJ...803...34B} luminosity functions.  Such a large cumulative number of these episodes could have a significant contribution to cosmic Reionization, especially as LAEs have higher values of \fion\ compared to continuum-selected samples \cite[e.g.][]{2015ApJ...809...89T,2018AA...614A..11M} and have number densities at $z > 6$ that are high enough that they can plausibly contribute significantly to Reionization \citep{2017AA...608A...6D}.  As stated above, understanding the distribution of \lya\ EWs and the relation between LAEs and the general galaxy population at these redshifts is critical, along with the duty cycle, to fully understand the potential contribution of these faint sources to Reionization.

A detailed discussion of the implications for Reionization is beyond the scope of this work.  Many studies combine the measured UV luminosity density $\rho_{\mathrm{UV}}$, derived from UV luminosity functions, with \xii\ and \fion\ in order to infer the ionizing emissivity $\dot{N}_{\mathrm{ion}}$ \cite[e.g.][]{2015ApJ...811..140B}.  As we show in Figure \ref{fig:xi_muv}, \xii\ can vary with the UV luminosity \cite[e.g.][]{2015MNRAS.451.2030D} and the dust content or metallicity of the galaxy \citep{2018ApJ...855...42S}.  Additionally, $\rho_{\mathrm{UV}}$ is not well-determined down to \muv\ $< -15$ as the majority of the observational constraints come from strong gravitational lensing where the associated systematic uncertainties are large \citep{2017ApJ...843..129B,2018MNRAS.479.5184A}.  Finally, the relationship between LAEs and the general population of galaxies, particularly at these extremely low luminosities, is unclear \cite[e.g.][]{2015MNRAS.446..566M}.

While these observations are the first step in understanding the ionizing photon production efficiency in such extreme systems, our method necessarily determines the average properties: individual objects could differ significantly.  Similar high-EW LAEs at $z < 4$ could be studied from the ground in the near-IR with detections of \hb\ in extremely long integrations with current instruments, or with future near-IR capabilities provided by the next generation of \textit{Extremely Large Telescopes}.  With the advent of new space-based facilities such as JWST, however, these measurements can be done out to higher redshifts using the brighter \ha\ emission line.  Combined with deep rest-UV imaging and spectroscopy to potentially measure \fion, we will obtain a more complete picture of the production and escape of ionizing photons from the abundant low-luminosity population of galaxies in the early Universe.

\section*{Acknowledgements}
We would like to thank the anonymous referee for a thoughtful report and suggestions that have improved this manuscript.  We are also grateful to everyone involved in the \textit{Spitzer} Space Telescope mission and everyone at the Spitzer Science Center: we are truly fortunate to have been able to use data from this facility.  JB acknowledges support by FCT/MCTES through national funds by this grant UID/FIS/04434/2019 and through the Investigador FCT Contract No. IF/01654/2014/CP1215/CT0003.  SC gratefully acknowledges support from Swiss National Science Foundation grant PP00P2$\_$163824.  We would also like to thank Mauro Stefanon for his assistance with de-blending the IRAC photometry, Pieter van Dokkum for a number of useful suggestions, and Daniel Schaerer for information regarding the stellar population models.

\bibliographystyle{mnras} 
\bibliography{muse}

\bsp	
\label{lastpage}
\end{document}